\documentclass{pasj00}
\draft
\usepackage{color}
\usepackage{rotating}
\usepackage{lscape}
\begin{document}
\SetRunningHead{Author(s) in page-head}{Running Head}
%\Received{}%{yyyy/mm/dd}
%\Accepted{}%{yyyy/mm/dd}
%\Published{}%{yyyy/mm/dd}
\title{Stars at the Tip of Peculiar Elephant Trunk-Like Clouds in IC 1848E: 
A Possible Third Mechanism of Triggered Star Formation}
%%% begin:list of authors
% Do NOT capitalize all letters in "textsc".

\author{Neelam {\textsc Chauhan},\altaffilmark{1}
        Katsuo {\textsc Ogura},\altaffilmark{2}
        Anil K. {\textsc Pandey},\altaffilmark{1}
        Manash R. \textsc{Samal},\altaffilmark{1}
        and
        Bhuwan C. {\textsc Bhatt}\altaffilmark{3}
        }
\altaffiltext{1}{Aryabhatta Research Institute of Observational Sciences (ARIES), 
Nainital 263 129, India}
\email{neelam@aries.res.in}
\altaffiltext{2}{Kokugakuin University, Higashi, Shibuya-ku, Tokyo 150-8440}
\email{ogura@kokugakuin.ac.jp}
\altaffiltext{3}{CREST, Indian Institute of Astrophysics, Hosakote 562 114, India}
%--------------------
%Johann KEPLER,1 Galileo GALILEI,1,2 and Isac NEWTON 
%1Institute 1
%2Institute 2

%\author{Neelam \textsc{Chauhan} 
%\affil{Aryabhatta Research Institute of Observational 
%Sciences (ARIES), Nainital 263 129, India}
%\email{neelam@aries.res.in}
%\author{Katsuo \textsc{Ogura}}
%\affil{Kokugakuin University, Higashi, Shibuya-ku, 150-8440 Tokyo}\email{ogura@kokugakuin.ca.jp}
%\author{Anil K. {\sc Pandey}}\affil{Aryabhatta Research Institute of Observational 
%Sciences (ARIES), Nainital 263 129, India}\email{pandey@aries.res.in}
%\author{Manas R. {\sc Samal}}\affil{Aryabhatta Research Institute of Observational 
%Sciences (ARIES), Nainital 263 129, India}\email{manash@aries.res.in}\and\author{Bhuwan C. {\sc Bhatt}}
%\affil{Indian Institute of Astrophysics, India}\email{ccccc@xxx.xxx.xxx}
%%% end:list of authors
%%% Please use the following style in case that sorting by 
%%% affilation is impossible.
% \author{%%   D-Firstname \textsc{D-Familyname}\altaffilmark{1} %   E-Firstname \textsc{E-Familyname}\altaffilmark{1,2} %   and%   F-Firstname \textsc{F-Familyname}\altaffilmark{2}} % \altaffiltext{1}{Address of Institute} % \email{ddddd@xxx.xxx.xx.xx} % \email{eeeee@xxx.xxx.xx.xx} % \altaffiltext{2}{Address of Institute} %% `\KeyWords{}' always has to be placed before `\maketitle'.
 %Do NOT move this preamble from here! 
\maketitle 

\begin{abstract}
The HII region IC 1848 harbors a lot of intricate elephant trunk-like 
structures that look morphologically different from usual bright-rimmed 
clouds (BRCs). Of particular interest is a concentration of thin and long 
elephant trunk-like structures in the southeastern part of IC 1848E. Some 
of them have an apparently associated star (or two stars) at their very tip. 
We conducted $VI_{c}$ photometry of several of these stars. Their positions 
on the $V/(V-I_{c})$ color-magnitude diagram as well as the physical 
parameters obtained by SED fittings indicate that they are low-mass 
pre-main-sequence stars having ages of mostly one {\it Myr} or less. This 
strongly suggests that they formed from elongated, elephant trunk-like 
clouds. We presume that such elephant trunk-like structures are genetically 
different from BRCs, on the basis of the differences in morphology, size 
distributions, and the ages of the associated young stars. We suspect that 
those clouds have been caused by hydrodynamical instability of the 
ionization/shock front of the expanding HII region. Similar structures often 
show up in recent numerical simulations of the evolution of HII regions. We 
further hypothesize that this mechanism makes a third mode of triggered star 
formation associated with HII regions, in addition to the two known mechanisms, 
i.e., {\it collect-and-collapse} of the shell accumulated around an expanding 
HII region and {\it radiation-driven implosion} of BRCs originated from 
pre-existing cloud clumps. 
\end{abstract}

\section{Introduction}

Recent high-resolution images of many HII regions taken with the {\it Hubble 
Space Telescope\/} and the {\it Spitzer Space Telescope} show very complicated 
structures inside them. One of such HII regions is IC 1848 (= W5). See, 
e.g., Figure 4 of Koenig et al. (2008), where we find a wealth of intricate 
structures inside/on its boundaries. Some of them are bright-rimmed clouds 
(BRCs) cataloged in Sugitani, Fukui, and Ogura (1991). But others are 
morphologically much different from usual BRCs, suggesting that they are 
genetically different from BRCs. We discuss this point in Sect. 4, but 
briefly, we suspect that, whereas BRCs mostly originate from pre-existing 
cloud clumps left-over in evolved HII regions, some of them may have resulted 
from the hydrodynamical instability of the ionization front of the expanding 
HII region. Of particular interest is a concentration of thin and long 
elephant trunk-like structures (hereafter abbreviated as {\it ETLS\/}s) in 
the southeastern part of IC 1848E. Figure 1 is a contrast-enhanced 
pseudo-color image of part of IC 1848E taken by the {\it Spitzer Space 
Telescope\/} (blue : 3.6$\mu$m, green : 8.0$\mu$m, red : 24.0$\mu$m). Note 
that all these {\it ETLS\/}s point to HD 18326, the exciting O star of 
IC 1848E. Very interestingly, some of them have a star/a few stars at their 
very tip, as marked in figure 1. This led us to suspect that they gave 
birth to these stars under the compressing effects of HII gas. Zavagno 
et al. (2007) found two intrusions of similar morphology with a star at 
their tip in RCW 120 (see their Fig. 12). We further suspect that the 
hydrodynamical instability of the ionization fronts creating {\it ETLS\/}s 
makes a third mechanism of triggered star formation associated with HII 
regions, in addition to the {\it collect-and-collapse} process of the 
shell accumulated around an expanding HII region and {\it radiation-driven 
implosion} of BRCs.

In order to examine the pre-main-sequence (PMS) nature of the stars located 
at the tip of the {\it ETLS\/}s, we carried out $VI_{c}$ photometry of these 
stars and constructed a $V/V-I_{c}$ color-magnitude diagram (CMD). We also 
used near-infrared (NIR) data from the {\it Two Micron All Sky Survey\/} 
(2MASS) to construct a NIR color-color diagram as well as mid-infrared 
(MIR) data from the {\it Spitzer Space Telescope} to make spectral energy 
distribution (SED) curves.

%%%%%%%%%%%%%%%%%%%%%%%%%%%%%%%%%Figure 1%%%%%%%%%%%%%%%%%%%%%%%%%%%%%%%%
\begin{figure}
  \begin{center}
\includegraphics[scale = 1.5, trim = 0 0 0 0, clip]{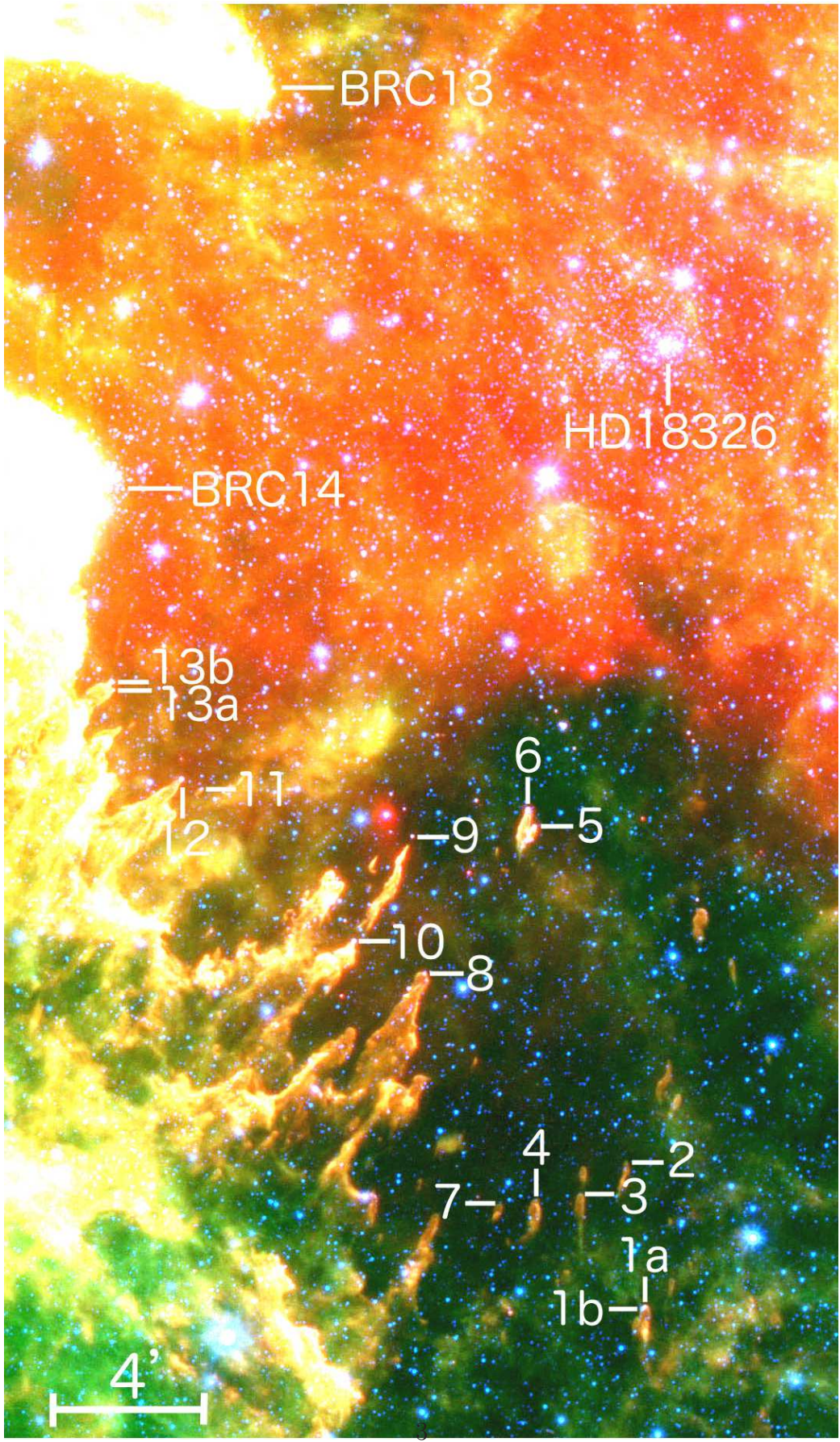}
%   \FigureFile(140mm,110mm){Fig1.eps}     %%% \FigureFile(width,height){filename}   
\end{center}
\caption{Contrast-enhanced {\it Spitzer} pseudo-color image of part of 
IC 1848E taken from the NASA {\it Spitzer Space Telescope} website. Stars at the 
tip of elephant trunk-like structures are marked together with two 
bright-rimmed clouds and the exciting star of IC 1848E. The scale is shown. 
North is up, east to the left.}
\label{fig1}
\end{figure}
%%%%%%%%%%%%%%%%%%%%%%%%%%%%%%%%%%%%%%%%%%%%%%%%%%%%%%%%%%%%%%%%%%%%%%%%% 

\section{Target Selection, Observations and Data Reduction}

BRCs are small clouds apparently with both width and length of several 
arcminutes, corresponding to the physical size of a few {\it parsecs} typically (see 
Sugitani et al. 1991, Sugitani \& Ogura 1994). In these papers BRCs are 
morphologically classified into {\it types A, B} and {\it C} according to their 
length-to-width ratios with {\it type C} being the most elongated. But many of the 
{\it ETLS\/}s found in figure 1 are much more elongated and of far smaller 
widths (typically one tenth of {\it pc}) than most of the {\it type C} BRCs 
in Sugitani et al. (1991) and Sugitani and Ogura (1994). We searched for such 
peculiar {\it ETLS\/}s that have a star/stars at their tip on the {\it Spitzer} 
3.6 $\mu$m, 4.5 $\mu$m, and 8.0 $\mu$m images. Table 1 gives the results, 
listing such stars with running numbers identified in figure 1, the coordinates 
and some remarks. We refer to these stars as {\it ETLS} stars. Some of them are 
listed in Koenig et al. (2008, their Table 4).  
%%%%%%%%%%%%%%%%%%%%%%%%%%%%%%%%%%%Table 1%%%%%%%%%%%%%%%%%%%%%%%%%%%%%%%%%%%
\begin{table*}
\caption{Stars at the tip of elephant trunk-like structures.} 
\begin{tabular}{|p{.7in}|p{.7in}|p{.9in}|p{2.3in}|}
\hline
Star ID & $\alpha_{(2000)}$ & $\delta_{(2000)}$ & Identification \& remarks \\
          & (h:m:s)          & ($\circ$ : $^\prime$ : $^{\prime\prime}$)           &                          \\
\hline
1a  & 02:59:18.08 & +60:08:37.7 &  \\
1b  & 02:59:18.61 & +60:08:34.5 & brighter than 1a \\
2   & 02:59:23.28 & +60:12:22.9 &  \\
3   & 02:59:32.93 & +60:11:32.1 &  \\
4   & 02:59:42.33 & +60:11:18.4 &  \\
5   & 02:59:46.25 & +60:21:09.7 & K14 \\
6   & 02:59:47.75 & +60:21:36.6 &  \\
7   & 02:59:49.66 & +60:11:13.1 &  \\
8   & 03:00:08.01 & +60:17:12.1 &  \\
9   & 03:00:12.21 & +60:20:45.4 &  \\
10  & 03:00:23.54 & +60:17:55.1 & K15 \\
11  & 03:00:57.52 & +60:21:44.0 &  \\
12  & 03:01:01.97 & +60:21:57.7 & K16 \\
13a & 03:01:17.42 & +60:24:13.4 &  \\
13b & 03:01:17.55 & +60:24:25.0 & K17? \\
%13c & 03:01:18.05 & +60:24:28.8 & bright in the opical \\
\hline
\end{tabular}
\\
\\
Note -- K numbers are identifications from Table 4 of Koenig et al. (2008). 
The coordinates of K17 are between those of 13a and 13b. There is another 
star $5^{\prime\prime}.4$ NE of 13b. It is relatively bright in the optical, 
but presumably a field star unrelated to the {\it ETLS} in view of its 
positions on the $V/(V-I_c)$ color-magnitude diagram and ${(J - H)/(H - K)}$ 
color-color diagram. 
\label{tab1}
\end{table*} 
%%%%%%%%%%%%%%%%%%%%%%%%%%%%%%%%%%%%%%%%%%%%%%%%%%%%%%%%%%%%%%%%%%%%%%%%%%%%

For the {\it ETLS} stars that are visible in the {\it DSS 2 red} image, we 
carried out photometric observations in the $V$ and $I_{c}$ bands using 
{\it Himalaya Faint Object Spectrograph Camera} (HFOSC) in the imaging mode 
mounted on the 2.0-m {\it Himalayan Chandra Telescope} (HCT) of the Indian 
Astronomical Observatory (IAO), Hanle, India on 2009 November 22, 23, and 24. 
HFOSC is equipped with a 2048 $\times$ 2048 pixel$^2$ CCD camera. The details 
of the site, HCT and HFOSC can be found at the HCT website 
(http://www.crest.ernet.in). The sky at the time of observations was 
photometric with a seeing size (FWHM) of $\sim$$1^{\prime\prime}.5$. A number 
of bias and twilight flat frames were also taken during the observing runs. 
The log of the HCT observations is tabulated in table \ref{tab2}. 

%%%%%%%%%%%%%%%%%%%%%%%%%%%%%%%%%%%Table 2%%%%%%%%%%%%%%%%%%%%%%%%%%%%%%%%%%
\begin{table*}
\caption{Log of observations.}
\begin{tabular}{|p{.7in}|p{.9in}|p{2.3in}|p{1.4in}|}
\hline
$\alpha_{(2000)}$ & $\delta_{(2000)}$ & Filter \& exposure (sec) $\times$ no. 
of frames& Date of observations\\
(h:m:s)           & ($\circ$ : $^\prime$ : $^{\prime\prime}$)           &                & (yr-mm-dd)\\
\hline
02:59:18.6 &+60:08:34& V:600$\times$3; I:200$\times$3 & 2009-11-22\\
03:20:07.0 &+60:18:47& V:600$\times$3; I:200$\times$3 & 2009-11-23\\
03:00:44.7 &+60:20:45& V:600$\times$2; I:200$\times$3 & 2009-11-24\\
\hline
\end{tabular}
\label{tab2}
\end{table*} 

%%%%%%%%%%%%%%%%%%%%%%%%%%%%%%%%%%%%%%%%%%%%%%%%%%%%%%%%%%%%%%%%%%%%%%%%%%% 
%%%%%%%%%%%%%%%%%%%%%%%%%%%%%%%%%%Table 3%%%%%%%%%%%%%%%%%%%%%%%%%%%%%%%%%%%
\begin{table}
\tiny
\caption{Photometric data for the {\it ETLS} stars.}
%\begin{sideways}
%\begin{minipage}{240mm}
\begin{tabular}{p{.05in}p{.48in}p{.48in}p{.48in}p{.48in}p{.48in}p{.55in}p{.55in}p{.55in}p{.55in}p{.5in}p{.2in}p{.02in}}
%\begin{tabular}{lllllllllllll} 
\hline
%Star &$V \pm \Delta $&$I_c \pm \Delta $&$J \pm \Delta $&$H \pm \Delta $&$K_s \pm \Delta $&$[3.6] \pm \Delta$&$[4.5] \pm \Delta$&$[5.8] \pm \Delta$&$[8.0] \pm \Delta$&$[24] \pm \Delta$&class$^{\dagger}$ \\
%ID &  $V (mag)$   &$I_c (mag)$    & $J (mag)$ & $H (mag)$ &$K_s (mag)$  &$[3.6] (mag)$  &$[4.5] (mag)$   &$[5.8] (mag)$&$[8.0] (mag)$&$[24] (mag)$& \\
Star &$V \pm \Delta V$&$I_c \pm \Delta I_c$&$J \pm \Delta J$&$H \pm \Delta H$&$K_s \pm \Delta K_s$&$[3.6] \pm \Delta [3.6]$&$[4.5] \pm \Delta [4.5]$&$[5.8] \pm \Delta [5.8]$&$[8.0] \pm \Delta [8.0]$&$[24] \pm \Delta [24]$&class$^{\dagger}$ \\
ID &  $(mag)$   &$(mag)$    & $(mag)$ & $(mag)$ &$(mag)$  &$(mag)$  &$(mag)$   &$(mag)$&$(mag)$&$(mag)$& \\

%&&&&&&&&&&&et al.\\
% &&&&&&&&&&&2008) \\
\hline
1b &$20.73 \pm 0.08$&$17.59 \pm 0.02$ & $ 14.63 \pm 0.04 $& $12.72 \pm 0.03 $& $11.78 \pm 0.03 $& $11.04 \pm 0.01 $& $10.92 \pm 0.01 $& $10.74 \pm 0.02 $& $10.56 \pm 0.06 $&-& III \\
5$^*$ &$19.49 \pm 0.02$&$17.25 \pm 0.01$& $ 16.14 \pm 0.11 $& $15.22 \pm 0.11 $& $14.38 \pm  0.1 $& $12.65 \pm 0.01 $& $ 11.9 \pm 0.01 $& $11.04 \pm 0.01 $& $  9.9 \pm 0.03 $& $ 6.68 \pm 0.16$& I   \\ 
6 &$20.13 \pm 0.02$&$17.35 \pm 0.01$&$15.41 \pm 0.08 $&$ 14.07 \pm 0.06 $&$12.96 \pm 0.04 $&$ 11.93 \pm 0.01 $&$ 11.27 \pm 0.01$ &$ 10.59 \pm 0.01 $&$  9.63 \pm 0.01$ &$ 6.11 \pm 0.12$& II  \\
9$^*$ &$19.81 \pm 0.02$&$16.93 \pm 0.01 $ & $ 15.08 \pm 0.05 $& $14.08 \pm 0.05 $& $ 13.5 \pm 0.04 $& $12.85 \pm 0.01 $& $12.41 \pm 0.01 $& $ 11.9 \pm 0.02 $& $10.99 \pm 0.02 $& $ 8.08 \pm 0.06$& II  \\
10$^*$ &$17.4 \pm 0.01$&$15.05 \pm 0.01$& $ 13.38 \pm 0.03 $& $12.48 \pm 0.03 $& $11.96 \pm 0.03 $& $11.29 \pm 0.01 $& $10.91 \pm 0.01 $& $10.48 \pm 0.01 $& $  9.9 \pm 0.01 $& $ 7.16 \pm 0.14$& II \\ 
12 &$18.66 \pm 0.01$&$16.03 \pm 0.01$& $ 14.31 \pm 0.03 $& $13.24 \pm 0.03 $& $12.64 \pm 0.03 $& $11.59 \pm 0.01 $& $11.08 \pm 0.01 $& $10.56 \pm 0.01 $& $ 9.73 \pm 0.01 $& $ 7.08 \pm 0.08$& II  \\ 
13b &$22.08 \pm 0.09$&$18.19 \pm 0.01$&$16.09 \pm 0.11$ &$14.84 \pm 0.08$&$14.14 \pm 0.08$& $13.01 \pm 0.01 $& $12.61 \pm 0.01 $& $12.06 \pm 0.02 $& $11.12 \pm 0.05 $& -& II   \\ 
%13c &$22.08 \pm 0.09$&$18.19 \pm 0.01$&$14.87 \pm 0.06$ &$14.29 \pm 0.08$&$14.05 \pm 0.06$& -& - & -& -& -& -   \\ 
\hline
\end{tabular}
\\
\\
$^*$ $V$ and $I_c$ magnitudes are averages of those obtained on different nights.\\
$^{\dagger}$ Koenig et al. (2008).
\label{tab3}
\end{table}
%%%%%%%%%%%%%%%%%%%%%%%%%%%%%%%%%%%%%%%%%%%%%%%%%%%%%%%%%%%%%%%%%%%%%%%%%%%%%%%%%%%%

The data analyses were carried out at ARIES, Nainital, India. The initial 
processing of the data frames was done using various tasks available under 
the IRAF data reduction software package. Photometric measurements of 
the {\it ETLS} stars were performed by using DAOPHOT II software package 
(Stetson 1987). A point spread function (PSF) was obtained for each frame 
using several uncontaminated stars. The results of the measurements were 
transformed to the standard system by using the secondary standards taken 
from Chauhan et al. (2011). The photometric accuracies depend on the 
brightness of the stars, and the typical DAOPHOT errors in the $V$ and $I_c$ 
bands at $V$ $\sim$ 18 are smaller than 0.01 mag. Near the limiting magnitude 
of $V$ $\sim$ 22 they increase to 0.1 and 0.02 mag in the $V$ and $I_c$ bands, 
respectively. 

Since young stellar objects (YSOs) often show NIR/MIR excesses caused by 
circumstellar disks, NIR/MIR photometric data are very important to know 
their nature and evolutionary status. $JHK_s$ data for the {\it ETLS} 
stars have been obtained from the 2MASS Point Source Catalog (PSC) (Cutri 
et al. 2003). 
%The JHKs data were transformed from the 2MASS system to the %CIT system using 
%the relations given in the 2MASS website. 
%In order to help elucidate the evolutionary stages of the {\it ETLS} stars 
%detected by  
Also we tried to collect {\it Infrared Array Camera} (IRAC) 3.6 $\mu$m, 4.5 
$\mu$m, 5.6 $\mu$m, and 5.8 $\mu$m data and {\it Multiband Imaging Photometer 
for Spitzer} (MIPS) 24 $\mu$m photometry for them from Koenig et al. (2008)'s 
list of stars in the W5 region. We searched for the 2MASS and {\it Spitzer} MIR 
counterparts of the {\it ETLS} stars and identified them using a search radius 
of $1.^{\prime\prime}2$. The photometric data for the stars are given in table 
\ref{tab3}.  

\section {Results}
\subsection{NIR Color-color Diagram}

Figure \ref{fig2}a (left) shows the ${(J - H)/(H - K)}$ NIR color-color diagram (CCD) 
for the {\it ETLS} stars identified in the 2MASS PSC catalogue. The solid and 
long-dashed curves represent the unreddened main sequence and giant branches 
(Bessell $\&$ Brett 1988), respectively. The dotted line indicates the loci of 
intrinsic classical T Tauri stars (CTTSs) (Meyer et al. 1997). The parallel 
dashed lines are reddening vectors drawn from the tip (spectral type M4) 
of the giant branch (``upper reddening line''), from the base (spectral type 
A0) of the main sequence branch (``middle reddening line'') and from the tip 
of the intrinsic CTTS line (``lower reddening line''). The extinction ratios 
$A_J/A_V = 0.265, A_H/A_V = 0.155$, and $A_K/A_V=0.090$ have been adopted from 
Cohen et al. (1981). All of the star positions and lines are in the CIT system. 
We classify the NIR CCD into three zones (`F', `T', and `P') to study the nature 
of the sources (for details see Ojha et al. 2004a, b). The `F' sources are 
located between the upper and middle reddening lines and are considered to be 
either main-sequence stars or weak-line T Tauri stars or CTTSs with small NIR 
excesses. `T' sources are located between the middle and lower reddening lines 
and are considered to be mostly CTTSs/Class II objects with large NIR excesses. 
The sources in the `P' region are most likely Class I stars (protostar-like 
objects), surrounded by an envelope. In figure \ref{fig2}a {\it ETLS} stars 
having a 2MASS counterpart, are plotted with different symbols according to 
the classifications by Koenig et al. (2008) based on the {\it Spitzer} data. 
Sources of class I, class II, and class III are shown by filled circles, by 
triangles, and by open circles, respectively. We estimated $A_V$ for each star 
by tracing them back to the intrinsic CTTS line of Meyer et al. (1997) along 
the reddening vector (for details, see Ogura et al. 2007). The mean of the 
individual $A_V$ values turned out to be $A_V$ = $2.4 \pm 1.2$ mag or 
$E(V - I_c)$ = $0.96 \pm 0.48$ mag, which we use in further discussions.

\subsection{Optical Color-Magnitude Diagram}
%%%%%%%%%%%%%%%%%%%%%%%%%%%%%%%%%%%%Figure 2%%%%%%%%%%%%%%%%%%%%%%%%%%%%%%%%%%%%
\begin{figure*}
\centering
\includegraphics[scale = .5, trim = 5 5 5  5, clip]{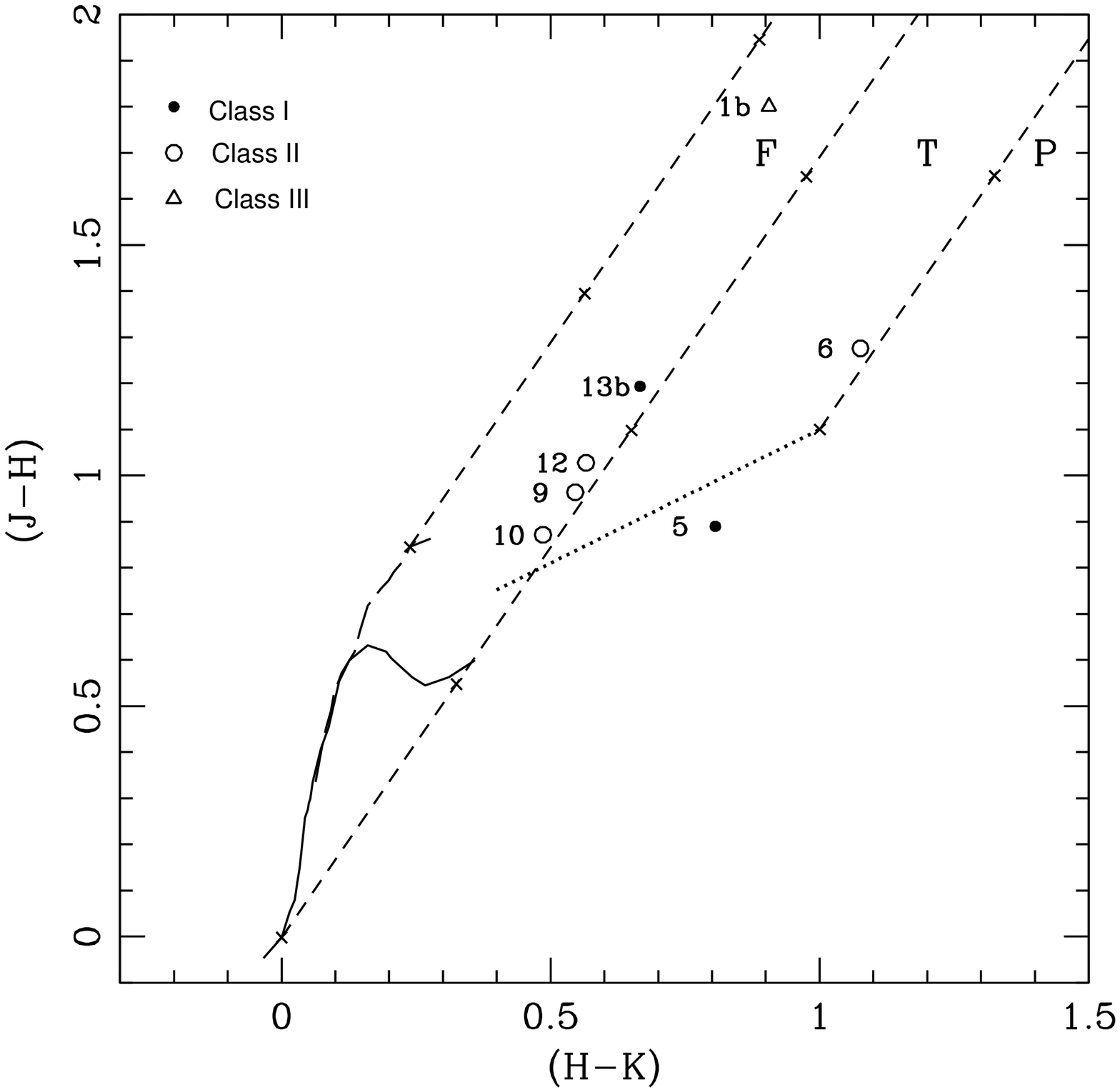}
\includegraphics[scale = .5, trim = 5 5 5  5, clip]{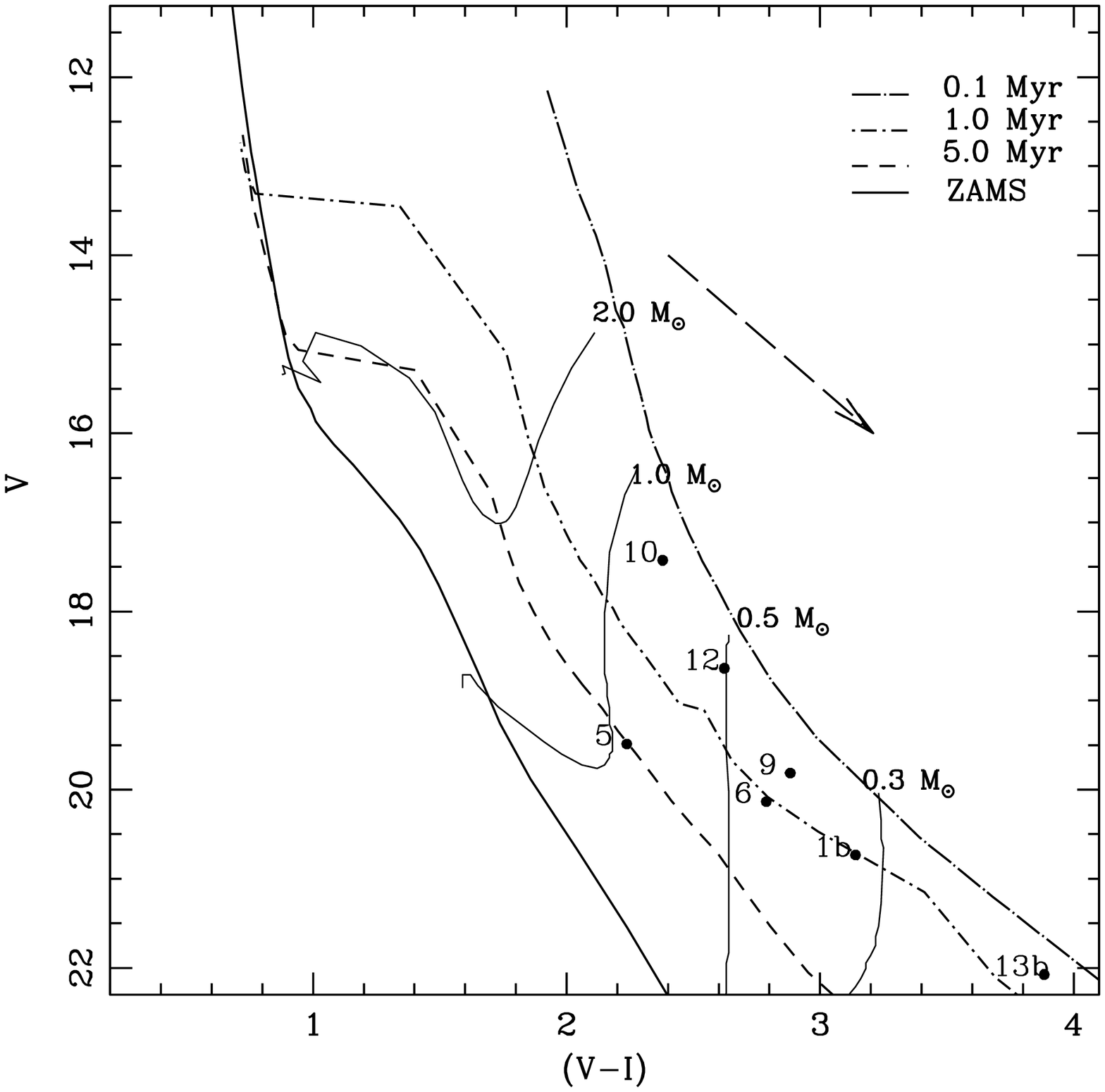}
\caption{(a, left) ${(J - H)/(H - K)}$ CCD for the {\it ETLS} stars. Their 
classifications are taken from Koenig et al. (2008), which are based on the 
$Spitzer$ data. The solid and thick dashed curves represent the unreddened 
main sequence and giant branches (Bessell \& Brett 1988), respectively. The 
dotted line indicates the loci of intrinsic CTTSs (Meyer et al. 1997). The 
parallel dashed lines are the reddening vectors drawn from the tip of the 
giant branch, from the base of the main sequence branch and from the tip of 
the intrinsic CTTS line with crosses representing a visual extinction of 
$A_V$ = 5 mag.  (b, right) $V/(V-I_c)$ CMD for the {\it ETLS} stars. The PMS 
isochrones and evolutionary tracks from Siess et al. (2000) are overplotted. 
The thick continuous line is the zero-age main sequence (ZAMS) from Girardi 
et al. (2002). The isochrones and evolutionary tracks are corrected for the 
distance 2.1 {\it kpc} and the mean reddening $E(V - I_c) = 0.96$ mag (see 
the text). The dashed arrow shows the reddening vector corresponding to 
$A_V$ = 2 mag.}
\label{fig2}
\end{figure*}
%%%%%%%%%%%%%%%%%%%%%%%%%%%%%%%%%%%%%%%%%%%%%%%%%%%%%%%%%%%%%%%%%%%%%%%%%%%%%

In figure \ref{fig2}b (right) we give the $V/(V-I_c)$ CMD of the {\it ETLS} stars 
listed in Table \ref{tab3}. The PMS isochrones and evolutionary tracks of 
Siess et al. (2000) as well as the zero-age main sequence of Girardi et al. 
(2008) are overlaid after being shifted to the distance modulus 13.5 mag 
(distance of 2.1 {\it kpc}; Chauhan et al. 2011) and the mean reddening 
of $E(V - I_c)$ = 0.96 mag. Note that the positions of the stars are not 
corrected for their reddening values. But the effect of the variable reddenings 
on the age estimation is small, because the reddening vector is nearly parallel 
to the $V/(V-I_c)$ PMS isochrones, as indicated in figure \ref{fig2}b. This 
CMD manifests that these sources are actually PMS stars having ages of 0.2 - 5 
{\it Myr} and masses of 0.1 - 1 {\it $M_\odot$}. We presume that these stars 
are physically related to the {\it ETLS\/}s because of their location at their 
very tip. However the possibility that some of them are field stars (foreground main-
sequence or background giant stars) can not be entirely rejected, since the 
southern part of IC 1848E is located at a very low $galactic$ latitude ({\it l} 
$\sim$ 1.5$^{\circ}$). 

\subsection{Spectral Energy Distribution Fitting}

To understand the nature and evolutionary status of the {\it ETLS} stars we 
re-construct their SEDs using the recently available grid of models and fitting 
tools of Robitaille et al. (2006, 2007). The models were computed using a 
Monte Carlo based radiation transfer codes (Whitney et al. 2003a, 2003b) assuming 
several combinations of a PMS central star, a flared accretion disk, a 
rotationally flattened infalling envelope and a bipolar cavity for a reasonably 
large parameter space. Interpreting SEDs using radiative transfer codes is 
subject to degeneracies, which spatially-resolved multiwavelength observations 
can overcome. The SED fitting tools fit these models to observational data 
points while assuming the distance and foreground reddening as being free parameters. 
For IC 1848 the distance in the literature varies from 1.9 to 2.3 {\it kpc} (Hillwig 
et al. 2006, Moffat 1972, Becker \& Fenkart 1971). Hence, we have taken the 
distance to be in the range of 1.9 to 2.3 {\it kpc}. Based on the NIR CCD 
(figure \ref{fig2}a), we assumed the visual absorption ($A_V$) ranges from 2 
to 10 mag for these sources. We set the uncertainties of the NIR and MIR 
flux estimates to be 10 to 15\%. We calculate a goodness-of-fit parameter, 
$\chi^2$, normalized by the number of data points {\it $N_{data}$} (9 or 10) 
used in the fitting. The evolutionary parameters of each source are 
determined by using the average of all the ``well-fitted'' models. The 
well-fitted models of each source are defined by \\

\begin{figure*}
\centering
\includegraphics[scale = 2.5, trim = 0 0 0 0, clip]{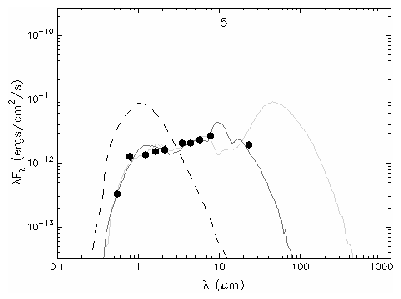}
\includegraphics[scale = 2.5, trim = 0 0 0 0, clip]{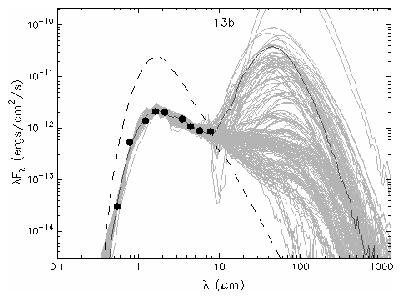}
\includegraphics[scale = 2.5, trim =0 0 0 10, clip]{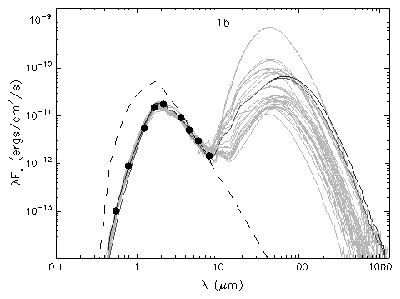}
\caption{SEDs for a Class I, Class II and Class III sources among the 
{\it ETLS} stars. The black lines represent the best fits, and the gray 
lines subsequent good fits. The dashed lines show the stellar photospheres 
corresponding to the central sources of the best-fit models. The filled 
circles are the input flux values. }
\label{sedfit}
\end{figure*}

   $\chi^2$ - ${\chi^2}_{min}$ $\le$ $2N_{data}$\\

where ${\chi^2}_{min}$ is the goodness-of-fit parameter of the best fit model. 
In table 4 we tabulate for each source the average parameters, such as the age, 
the interstellar extinction ($A_V$, which does not include the extinction due to 
the circumstellar disk or envelope), the mass of the star ($M_{star}$), the disk 
accretion rate ($\dot M_{disk}$), and the envelope accretion rate ($\dot M_{env}$). 
Here, it is worth mentioning that these are crude values, and should be 
considered to only be approximate in view of the underlying assumptions in the models. 
Also, the number of observational data points is limited in spite of many 
parameters involved. The stellar ages given in table \ref{sed} range from 0.2 to 
5 {\it Myr} again, although the results for individual stars differ from those  
derived from the $V/(V-I_c)$ CMD. Figure \ref{sedfit} shows three examples of the 
SED fitting.  

%%%%%%%%%%%%%%%%%%%%%%%%%%%%%%%%%%%Table 4%%%%%%%%%%%%%%%%%%%%%%%%%%%%%%%%%%
\begin{table}
\small
\caption{Physical parameters for the {\it ETLS} stars based on the SED fitting.}
\label{sed}
\begin{tabular}{ccccccccc} \hline
%.N.& Identification number& $\chi ^ 2$&$A_V$&    cluster &  & Field region\\
% (mag)& $r\le 3^\prime$ & $3^\prime <r \le6^\prime$&  \\
Star   & Age           & $ A_V$        &$ M_{star}$    &$M_{disk}$      &$\dot M_{disk}$        &$\dot M_{env}$          &${\chi^2}_{min}$& $N_{data}$\\
ID     & ({\it Myr})   & (mag)         &($M_\odot$)    &($M_\odot$)     &(10$^{-8}$$M_\odot$/yr)&(10$^{-6}$$M_\odot$/yr) &                &           \\
\hline                                                                                                                   
1b     & $0.2 \pm 0.1$ &  $3.8 \pm 0.8$&$1.8 \pm 1.2$  &$0.01 \pm 0.03$ &$11.0 \pm 11.0$        & $14.0 \pm 18.0$        & 5.36           &     9     \\ 
5      & $4.4 \pm 3.1$ &  $ 3.1\pm 0.6$&$2.4 \pm 0.3$  &$0.02 \pm 0.02$ &$0.12 \pm 0.12$        & $1.3  \pm 1.7 $        & 25.85          &    10     \\
6      & $5.5 \pm 3.6$ &  $5.5 \pm 1.5$&$2.0 \pm 1.3$  &$0.01 \pm 0.02$ &$14.0 \pm 1.8$         & $1.1  \pm 11.0$        & 9.20           &    10     \\
9      & $2.8 \pm 2.1$ &  $4.3 \pm 0.8$&$1.6 \pm 0.5$  &$0.01 \pm 0.01$ &$0.5 \pm 0.5$          & $0.03 \pm 0.14$        & 5.06           &    10     \\
10     & $1.0 \pm 0.9$ &  $3.2 \pm 0.6$&$2.2 \pm 0.8$  &$0.01 \pm 0.01$ &$1.4 \pm 1.3$          & $0.57 \pm 1.2 $        & 2.15           &    10     \\
12     & $4.8 \pm 2.6$ &  $5.1 \pm 0.8$&$2.3 \pm 0.6$  &$0.01 \pm 0.01$ &$3.9 \pm 3.6$          & $0.1  \pm 0.4 $        & 5.31           &    10     \\
13b    & $2.3 \pm 2.3$ &  $5.4 \pm 1.0$&$2.1 \pm 0.9$  &$0.01 \pm 0.01$ &$1.6 \pm 1.5$          & $6.8 \pm 16.9$         & 7.44           &     9     \\
\hline
\end{tabular}
\end{table}
%%%%%%%%%%%%%%%%%%%%%%%%%%%%%%%%%%%%%%%%%%%%%%%%%%%%%%%%%%%%%%%%%%%%%%%%%%%%
%%%%%%%%%%%%%%%%%%%%%%%%%%%%%%%%%%%%Figure 3%%%%%%%%%%%%%%%%%%%%%%%%%%%%%%%%
%%%%%%%%%%%%%%%%%%%%%%%%%%%%%%%%%%%%%%%%%%%%%%%%%%%%%%%%%%%%%%%%%%%%%%%%%%

\section {Discussion}

\subsection {Origin of Bright-Rimmed Clouds and Hydrodynamical Instability of 
Ionization Fronts}

The effects of intense UV radiation from OB stars on star formation can be either 
constructive or destructive, depending on the situation. As for the mechanisms 
with which it works constructively, two have so far been proposed: 
{\it collect-and-collapse} and {\it radiation-driven implosion} (RDI). The 
former/latter is of larger/smaller size in space ($\sim$10 {\it pc}/$\sim$1 
{\it pc}) and of longer/shorter timescale ($\sim$a few {\it Myr}/$\sim$0.5 
{\it Myr}).

{\it Collect-and-collapse} was advocated by Elmegreen and Lada (1977) in their 
hypothesis of {\it Sequential Star Formation}. In this scenario, pressure-driven 
expansion of an HII region collects a dense shell between the ionization front 
(IF) and shock front (SF), which in due time becomes gravitationally unstable 
and collapses to form stars of the second generation including OB stars. Since 
then various analytic and numerical calculations were carried out under this 
scenario. However, it has never been convincingly confirmed for many years, 
until very recently when the Deharveng group (Deharveng et al. 2005; Pomar{\`e}s 
et al. 2009, and references therein) presented the first persuasive examples. 
This mechanism is probably viable in relatively uniform molecular clouds. 

RDI takes place in small molecular clouds, which are called ``{\it 
bright-rimmed clouds} (BRCs)'', ``{\it globules}'', ``{\it elephant trunks}'' 
and so forth. They are usually considered to be remnant cloud clumps left over 
in expanding HII regions. Detailed numerical calculations (e.g., Lefloch \& 
Lazareff 1994) showed that such clouds are compressed by the high pressure of 
the surrounding HII gas. Star formation in BRCs was suspected from early times 
(e.g., Wootten et al. 1983). Clear evidence for star formation in these clouds 
was provided by Sugitani, Fukui, and Ogura (1991) and Sugitani and Ogura (1994), 
who showed their association with IRAS point sources of low temperatures. Also, 
Sugitani, Tamura, and Ogura (1995) indicated that BRCs are often associated 
with a small star cluster, showing not only an asymmetric spatial distribution, 
but also a possible age gradient. This lead them to advocate the hypothesis 
of ``{\it small-scale sequential star formation}'', which has recently been 
verified quantitatively by {\it BVI$_{c}$} photometry by Ogura et al. (2007) 
and Chauhan et al. (2009, 2011). Detailed observations of physical properties 
of BRCs cataloged in Sugitani et al. (1991) and Sugitani and Ogura (1994) were 
made by the British group (Morgan et al. 2008, Urquhart, Morgan \& Thompson 
2009, and references therein) by means of sub-millimeter observations and radio 
continuum and CO/$^{13}$CO/C$^{18}$O line observations. They concluded that 
RDI is in progress in many (but not all) of these BRCs, and that relatively 
massive stars are being formed there, based on the high luminosity of the 
embedded sources (Urquhart et al. 2009). 

As for the origin of BRCs or elephant trunks, a Rayleigh-Taylor instability in  
expanding HII regions was proposed first (e.g., Spitzer 1954), but Pottasch 
(1958) pointed out disagreements between their morphology and the theoretical 
predictions. In mid-1960s Axford (1964) investigated the stability of weak 
D-type IFs, explicitly taking into account the effect of diffuse UV radiation 
caused by recombinations to the ground state of hydrogen atoms. He claimed 
that weak D-type IFs, which correspond to the major part of the evolution of 
HII regions, are stable against the growth of wavelengths larger than 0.2 
{\it pc}, so hydrodynamical instability could not be the origin of elephant 
trunks. Given the fact that radio observations showed the clumpiness of 
molecular clouds, BRCs or elephant trunks have since then been usually considered 
to be pre-existing cloud clumps left over in expanding HII regions. 
Sysoev (1997) re-examined the stability of D-type IFs analytically and showed 
that, contrary to the conclusion by Axford (1964), they are {\it not} stable 
even with the effect of the recombinations. This new result was confirmed by 
numerical simulations of Williams (2002). Prior to these studies, Giuliani 
(1979) investigated the stability of the combined systems of a D-type IF and a 
preceding SF and reached a similar conclusion that there is a new regime of 
instability (longer wavelengths which are similar to the widths of the above 
structures) that grows rapidly in an oscillatory manner (overstability). Vishniac 
(1983) generalized this instability including SN/wind bubble SFs, and it is now 
called as ``{\it thin shell instability}'' or ``{\it Vishniac instability}''. 
Its mechanism is simple, as shown, e.g., in figure 1 of Garc{\'i}a-Segura and Franco 
(1996). The presence of an IF exacerbates the growth of the instability. 

Thus, BRC- or globule-like structures seem to be also formed via hydrodynamical 
instability without pre-existing molecular clumps. This was clearly shown in 
numerical simulations (2-dimensional) of the evolution of HII regions by 
Garc{\'i}a-Segura and Franco (1996); such structures arise in all of their models 
as HII regions expand. In their 3-D simulations Whalen and Norman (2008) 
obtained very similar results to those of Garc{\'i}a-Segura and Franco (1996). 
Other recent numerical simulations of evolution of HII regions with turbulence 
(Mellema et al. 2006; Dale, Clark \& Bonell 2007; Gritschneder et al. 2009, 
2010) and without turbulence (Mizuta et al. 2006; Bisbas 
et al. 2009) all show the formation of BRC- or globule-like structures. But, 
since molecular clouds are very clumpy, it seems more likely that ordinary 
BRCs have their origin in pre-existing clumps. 

%%%%%%%%%%%%%%%%%%%%%%%%%%%%%%%%%%%%Figure 4%%%%%%%%%%%%%%%%%%%%%%%%%%%%%%%%
\begin{figure*}
\centering
\includegraphics[scale = .62, trim = 0 0 0 0, clip]{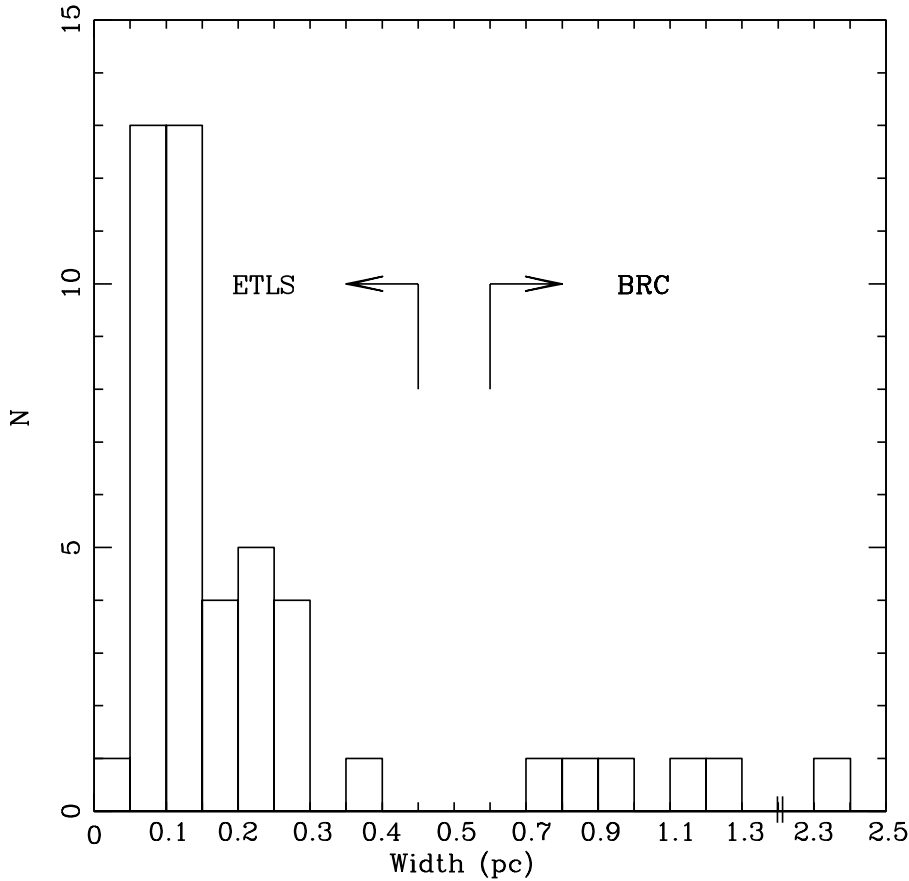}
\caption{Distribution of the widths of the head part of 41 {\it ETLS\/}s 
found in figure 1 as well as that of 6 BRCs listed in Ogura et al. (2002) 
in the whole IC 1848. Note that the abscissa scale of the right-hand half 
of the figure is two times smaller than that of the left-hand half.}
\label{hist}
\end{figure*}
%%%%%%%%%%%%%%%%%%%%%%%%%%%%%%%%%%%%%%%%%%%%%%%%%%%%%%%%%%%%%%%%%%%%%%%%%%

As for {\it ETLS\/}s, we suppose that their origin is different from that of 
ordinary BRCs and that they presumably originate from the above-mentioned hydrodynamical 
instability, based on the following three reasons. First, as mentioned already, 
the morphologies are very different; {\it ETLS\/}s are much thinner and more 
elongated than BRCs. Second, {\it ETLS\/}s and BRCs have different size 
distributions. Figure 4 shows the distribution of the widths of the head part 
of 41 {\it ETLS\/}s found in figure 1. That of BRCs is also shown for 
comparison; there are 6 BRCs listed in Ogura et al. (2002) in the whole IC 
1848, i.e., BRCs 11, 11NE, 11E, and 12 in IC 1848W, and BRCs 13 and 14 in 
IC 1848E. Note that the abscissa scale of the right-hand half of the figure 
is two times smaller than that of the left-hand half. Also, the number for 
the smallest bin may be affected by the incompleteness in picking up tiny 
{\it ETLS\/}s. The histogram shows a clear gap between the distributions of 
{\it ETLS\/}s and BRCs. Also, the combined size distribution of {\it ETLS\/}s 
and BRCs as well as that of the former, itself, does not exhibit any power laws, 
contrary to the well-known power-law core mass function (above a certain mass) 
(see, e.g., Sadavoy et al. 2010). There seems to be a peak at around 0.1 
{\it pc}. It might reflect the characteristic wavelength of the hydrodynamical 
instability in the IC 1848E HII region. The third reason is the fact that 
generally the {\it ETLS} stars are slightly {\it younger} than the stars 
associated with BRCs in IC 1848E. Figure 2b indicates the {\it ETLS} stars 
have ages of 0.2 - 1.0 {\it Myr} except for star No. 5. On the other hand, 
Chauhan et al. (2009) obtained 0.5 - 5 {\it Myr} and 0.1 - 3 {\it Myr} for 
majorities of the stars associated with BRCs 13 and 14, respectively (see 
their Figure 2). Chauhan et al. (2011) revisited these BRCs, and the results 
are 0.5 - 5 {\it Myr} and 0.3 - 5 {\it Myr}, respectively  (see their 
figure 9). From the very elongated morphology of the {\it ETLS\/}s one can 
imagine that they might be an older version of BRCs of the similar type, i.e., 
{\it type C} that formed from pre-existing clumps. But the above ages defy 
this conjecture.  

\subsection {Third Possible Mechanism of Triggered Star Formation}

On the basis of the result that the {\it ETLS} stars in IC 1848E are of the PMS 
nature having ages of 0.2 - 5 {\it Myr} and masses of 0.1 - 1 {\it $M_\odot$} , 
we consider that they formed under the compressing effects of the HII gas from 
these small clouds, which were created by a hydrodynamical instability of the 
expanding HII region. Thus, this process seems to make a third mode of triggered 
star formation associated with HII regions, in addition to {\it 
collect-and-collapse} and RDI. 

This new mechanism of triggered star formation is somewhat similar to the RDI 
in BRCs, but it differs in that the cloud was not pre-existing but formed from 
accumulated and then fragmented gas in the process of expansion of an HII region. 
In addition, we find only one star or at most a few stars at the tip of each 
{\it ETLS}, so the scale of star formation in each cloud is very small. However, the 
total product can be considerable because a large number of such structures can 
be formed in an HII region, as in IC 1848E. In our recent studies on BRC star 
formation we noticed many IR-excess stars scattered inside HII regions besides 
IC 1848E (see Fig. A3 of Chauhan et al. 2009). We suspect that some of these 
stars may have been formed by this mechanism. On the {\it Spitzer} IRAC images 
of the Carina Nebula Smith et al. (2010) also found a large number of scattered 
YSOs as well as many clouds morphologically similar to our {\it ETLS\/}s. Table 
5 summarizes the differences of this mechanism from {\it collect-and-collapse} 
and usual RDI. 

%%%%%%%%%%%%%%%%%%%%%%%%%%%%%%%%%%%Table 5%%%%%%%%%%%%%%%%%%%%%%%%%%%%%%%%%
\begin{table*}
\caption{Comparison of modes of triggered star formation.}
\begin{center}
\begin{tabular}{|p{1.2in}|p{1.0in}|p{.5in}|p{.9in}|p{.9in}|}
\hline
mode    & cloud  & scale & stars formed & timescale \\
\hline 
collect \& collapse  & accumulated & large & $>$ 300 & a few Myr \\
RDI & pre-existing & small & $<$ 100 & $<$ 1 Myr \\
HD instability & accumulated & small & $\le$ a few & $<$ 1 Myr \\
\hline
\end{tabular}
\label{tab5}
\end{center}
\end{table*}
%%%%%%%%%%%%%%%%%%%%%%%%%%%%%%%%%%%%%%%%%%%%%%%%%%%%%%%%%%%%%%%%%%%%%%%%%%%%

\section {Conclusions}

We paid attention to the numerous, elephant trunk-like clouds in IC 1848E and 
carried out $VI_{c}$ photometry of the optically visible stars located at the 
tip of several of them. Their positions on the $V/V-I_{c}$ CMD indicate that 
they are low-mass PMS stars of ages of mostly one {\it Myr} or less. The physical 
parameters derived for these stars by using the SED fitting tools indicate 
that they are largely Class I or Class II PMS sources. The PMS nature of these 
stars strongly suggests that they must have formed from these {\it ETLS\/}s. 
On the basis of the morphology, the size distributions, and the ages of the 
associated young stars we conclude that the {\it ETLS\/}s and BRCs have different 
origins, and suspect that the former are created by the hydrodynamical instability 
of the IF/SF of the expanding HII region. We further hypothesize that, in addition 
to the {\it collect-and-collapse} process and RDI, this mechanism makes a third 
mode of triggered star formation associated with HII regions. 

\bigskip
\section{Acknowledgement}
We are grateful to the anonymous referee for his/her useful comments that 
improved this paper. We thank the staff of IOA, Hanle and CREST, Hosakote for 
the assistance during the observations. NC is thankful to the fellowship 
granted by DST and CSIR, India. KO and AKP acknowledge JSPS, Japan and DST, 
India for the financial supports.

\end{document}